\documentstyle[aps,epsf,amsfonts]{revtex}
\begin{document}
\title{Correlations in doped Antiferromagnets}
\author{Moshe Havilio\footnote{havilio@pharaoh.technion.ac.il} 
and Assa Auerbach\footnote{assa@pharaoh.technion.ac.il}}
\address{Physics Department, Technion, Haifa 32000, Israel}
\date{\today}
\maketitle
\begin{abstract}
A comprehensive study of doped RVB states is performed. It reveals a   
 fundamental connection between superconductivity 
and  quantum spin fluctuations in underdoped cuprates :
 {\em  Cooper pair hopping} strongly reduces the local magnetization $m_0$.
This effect pertains to recent 
 muon spin rotation measurements 
in which $m_0$ varies weakly with hole doping in the
poorly conducting regime,  but drops precipitously above the onset
of superconductivity.
Gutzwiller mean field Approximation (GA) is found to agree with  numerical 
Monte Carlo calculation. GA shows for example that for a bond amplitude $u(r)=e^{-r/\xi}$,  spin spin correlations
decay exponentially with a correlation length $\propto e^{3\pi \xi^2/2}$.
Expectation value of the Heisenberg model is  found to be correlated with average loop density.

\end{abstract}
\section{Introduction}

When holes are introduced into the copper oxide planes of high T$_c$ Cuprates,
spin and charge correlations change dramatically. 
The  {\em local} magnetization $m_0$, measured by  $\mu$SR  
\cite{Niedermayer}  on e.g. $\mbox{La}_{2-x}\mbox{Sr}_x \mbox{CuO}_4$,
reveals a qualitative difference 
between the insulating and superconducting phases: $m_0$ is rather insensitive to doping in the poorly conducting
regime 
$0\le x\le 0.06$, but drops precipitously above the onset  of  superconductivity at $x > 0.06$, 
becoming undetectable at optimal doping  $x\approx 0.15$. Theoretically, 
holes can cause {\em dilution} and 
{\em frustration} in  the Heisenberg antiferromagnet, which create spin textures: 
either random (``spin glass'') or with ordering wavevector away from $(\pi,\pi)$ 
(sometimes called ``stripes'')\cite{IC}. However, the apparent reduction of local magnetization 
by the onset of superonductivity, is a novel and poorly understood effect. 
Theory must go beyond
purely magnetic models, and
involve the superconducting degrees
of freedom.  

We find that this problem is amenable to a variational approach, using  hole-doped 
Resonating Valence Bonds (RVB)  states. The RVB states were originally suggested by 
Anderson to describe the spin and charge
correlations in the  high $T_c$ Cuprates\cite{PWA}. They are excellent trial wave functions for the  
doped Mott insulators, with large Hubbard repulsion $U$ since:\newline
(i) Configurations with doubly occupied sites are excluded.\newline
(ii)  Marshall's sign criterion for the magnetic energy\cite{Marshall} is satisfied,
and Heisenberg ground state energy and antiferromagnetism at  zero doping is accurately recovered\cite{LDA,japan}.

The hole-doped RVB state is a new class of variational states, in which
spin and charge correlations are parameterized independently,
without explicit spin nor gauge symmetry breaking. Such parameterization allows  states with
magnetic  and independently d or s-wave superconducting (off-diagonal) order or disorder, thus
permit an {\em unbiased} determination of 
ground state spin and charge correlations appropriate for the Cuprates.
These are important advantages over commonly
used Spin Density Wave, Hartree-Fock and BCS wavefunctions.

A comprehensive  study of the state is performed using Monte Carlo and mean field calculations.
Phenomenological low energy effective Hamiltonian is proposed,
with two major components:
Heisenberg interaction for spins, 
and single or Cooper pair hopping kinetic energy for fermion holes.

Regarding this model our key results are:\newline
(i)  For the magnetic energy alone,  the local magnetization $m_0$  is 
{\em weakly dependent} on doping concentration. This holds
independently of inter-hole correlations for either
randomly localized or extended states.\newline
(ii)  In contrast to (i), $m_0$ is strongly reduced by the kinetic
energy of {\em Cooper pair hopping}, which correlates the reduction of $m_0$
with the  rise of
superconducting stiffness, and  hence\cite{emkivsfs} the
transition temperature $T_c$.

These   results agree with the experimentally reported correlation between $m_0$ and  $T_c$\cite{Niedermayer}. 
This relation appears to be  {\em weakly} dependent on the precise
hole density.

We also find that RVB states have the following  properties:\newline
(i) The magnetic energy is correlated with the average loop density:\newline
$\Gamma=L^2m_0^2/(\mbox{average radius of gyration of a loop})^2$,
 where $L$ is the linear size off the lattice.\newline
(ii) The Gutzwiller mean field Approximation (GA) for magnetic correlations is in good agreement with the 
Monte Carlo results.\newline
(iii) Long range magnetic correlations in RVB states are extremely sensitive to changes in the singlet bond amplitude $u$.
For example with $u(r)= \exp\left(-r / \xi \right)$  the spin-spin correlation function 
decays exponentially with correlation length $\xi_{ex} \propto \exp{\left((1-x ) \frac{3 \pi}{2} \xi^2\right)}$,
where $x$ is the hole concentration.

The paper is organized as followed:
Sec.\ref{srv} introduces the hole-doped RVB state, and discusses the numerical procedure.
Sec.\ref{uuuu} defines our variational parameters.
Sec.\ref{orpar} deals with  the antiferromagnetic and superconducting order parameters.
Sec.\ref{efh} deals with the components of the effective Hamiltonian.
Sec.\ref{tcm0} correlates between superconducting T$_c$ and local magnetization.
Sec.\ref{diss} is a summary and discussion.

The paper has 3 appendices. 
App. \ref{ferpart} reduces the hole part of the doped RVB to  a numerically convenient format. 
App. \ref{esr} derives expressions for expectation values. Particularly, an alternative procedure to calculate
magnetic correlation is derived and used to check the computer program. 
In App.\ref{largen}, the GA is performed analytically.

\section{The Hole doped Resonating Valence Bond states}
\label{srv}

A Valence bond (VB) state is 
\begin{equation}
|\alpha \rangle=\prod_{(i,j) \in \alpha}(a^{\dagger}_i b^{\dagger}_j - b^{\dagger}_ia^{\dagger}_j)|0\rangle,
\label{vbs}
\end{equation}
where $\alpha $ is a pair covering of the lattice, $a^{\dagger}_i,\;b^{\dagger}_j $
are Schwinger bosons, and $i=1,\ldots L^2$ is a site index on a square lattice.

RVB states are superposition of VB states. We restrict the disscussion to 
\begin{equation}
|\Psi [u]\rangle=\sum_{\alpha}\prod_{(i,j)\in \alpha}u_{ij} 
(a^{\dagger}_i b^{\dagger}_j - b^{\dagger}_ia^{\dagger}_j)|0\rangle.
\label{rvb1}
\end{equation} 
where $u({\bf r}_{ij})\geq 0$ is a variational singlet bond  amplitude,
which connects sites of different  sublattices  $A$ and $B$ only.
This ensures Marshall's sign \cite{Marshall}.  

\begin{figure}
\centerline{ \epsfysize=3.0cm
\epsfbox{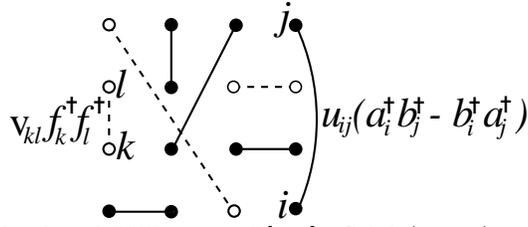}}
\caption[A bond configuration in the doped RVB states]
{A bond configuration in the doped RVB states $\Psi[u,v]$.
Solid (empty) circles represent spins (holes) with bond correlations $u_{ij}$ ($v_{kl}$). }
\label{rvbfig}  	 	 
\end{figure}

The  hole doped RVB state is defined by :
\begin{eqnarray}
 |\Psi [u,v;x]\rangle  &=&   
{\cal P}_G(x) |\bar{\psi}[u,v]\rangle \nonumber\\
  |\bar{\psi} [u,v]\rangle &\equiv&
\exp\left[
\sum_{i\in A,j\in B}\left( v_{ij} f^\dagger_i f^\dagger_j +u_{ij}
(a^{\dagger}_i b^{\dagger}_j - b^{\dagger}_ia^{\dagger}_j)\right)\right]|0\rangle
\label{states2}
\end{eqnarray}
where $f^\dagger_i$ are spinless hole fermions, $u_{ij}\geq 0$, and $v({\bf  r}_{ij})$ is an {\em independent}
hole bond parameter.
 The Gutzwiller projector  ${\cal P}_G(x)$ imposes two  constraints. 
A  constraint of no double occupancy:
\begin{equation}
n^i_a + n^i_b + n^i_f= 1 \;\;\;\; \forall  i,
\label{constrain}
\end{equation}
and a global constraint on the total number of holes:
\begin{equation}
\sum_i  n^i_f=x L^2=N_h.
\label{mcon}
\end{equation}
 Due to ${\cal P}_G(x)$, $\Psi$ can be written as a sum over
bond configurations of singlets and hole pairs which cover the lattice as depicted in Fig. \ref{rvbfig}.

An overlap of two VB states, $\langle\alpha|\beta\rangle$, is expressed in terms of a directed loop covering 
of the lattice (DLC)  \cite{LDA,shafir,Book}, hence :
\begin{equation}
\langle\Psi[u]|\Psi[u]\rangle = \sum_{\Lambda}\Omega_{\Lambda} 
\label{n0}
\end{equation}
where $\Lambda$ is a  DLC,
\begin{equation}
\Omega_{\Lambda} \equiv  
\prod_{\lambda \in \Lambda}\left(2 \prod_{(i, j)\in \lambda}
u_{ij}\right)
\label{dlx}
\end{equation}
and $\lambda$  is a directed loop. 

With the results of App.\ref{ferpart}, the norm of the doped RVB state is
\begin{equation}
\langle \Psi [u,v;x] | \Psi [u,v;x]\rangle=
\sum_{\gamma, \Lambda(\gamma)}  W\left( \gamma, \Lambda(\gamma)\right),
\label{exnorm}
\end{equation}
where $\gamma$ is a {\em distinct} configuration of $N_h$ holes sites:
\begin{eqnarray*}
\gamma \equiv \left\{\left(i_k\in A, j_k\in B\right)\right\}_{k=1}^{\frac{N_h}{2}}:
 \forall k < k' \;\; i_k < i_{k'},\;  j_k <  j_{k'}, 
\end{eqnarray*}
$\Lambda_{\gamma}$ is a DLC which coveres the lattice but the hole sites,
\begin{equation}
W(\gamma, \Lambda_{\gamma})=\left\{ \begin{array}{ll}
  \rm{det}^2 V\! \left( \gamma \right) 
\Omega_{\Lambda_{\gamma}}& x>0  \\
 \Omega_{\Lambda}  & x=0 
\end{array} \right. 
\label{wwf}
\end{equation}
and $V$ is an $N_h/2\times N_h/2$ matrix with 
\begin{equation}
V\!\left(\gamma\right)_{kl} \equiv v_{i_k j_l}.
\label{calv}
\end{equation}

Expectation value of an operator $O$  is expressed as a weighted sum
\begin{eqnarray}
\langle O\rangle= \frac{1}{\langle \Psi | \Psi \rangle}
\sum_{\gamma, \Lambda_{\gamma}} W(\gamma, \Lambda _{\gamma})
O(\gamma, \Lambda_{\gamma}) \equiv \overline{O}
\label{ooooo}
\end{eqnarray}
where  $O(\gamma, \Lambda_{\gamma})$
is defined by  Eqs. (\ref{wwf})  and (\ref{ooooo}).

We use standard Metropolis algorithm\cite{mcbook} for the evaluation of  sum (\ref{ooooo}). 
The basic Monte Carlo step for updating the DLCs is the one used by Ref. \cite{LDA} : Choose at random a site and one of
its next-nearest neighbors  and exchange, 
with transition probability that satisfy detailed balance, 
the bonds connecting each
of them, either to the next site (forward-bond),  or the previous site  in their loops. 
In Ref. \cite{thesis} we show, that for $u_r>0\;\; \forall r$, 
these steps  are  {\em ergodic}, that is, any DLC can 
be reached from any other by a sequence of  Monte Carlo steps.

For the fermion holes our update scheme is a  simple generalization of the 
``inverse-update'' algorithm of 
Ceperley, Chester and Kalos \cite{Ceperley}. According to Eq. (\ref{calv}),
 changing a position of an $A$ ($B$) sublattice hole amounts to changing 
one row (column)  in the matrix $V$. 
In the our calculation boundary conditions are periodic.

\begin{figure}
\centerline{ \epsfysize = 9.0cm
\epsfbox{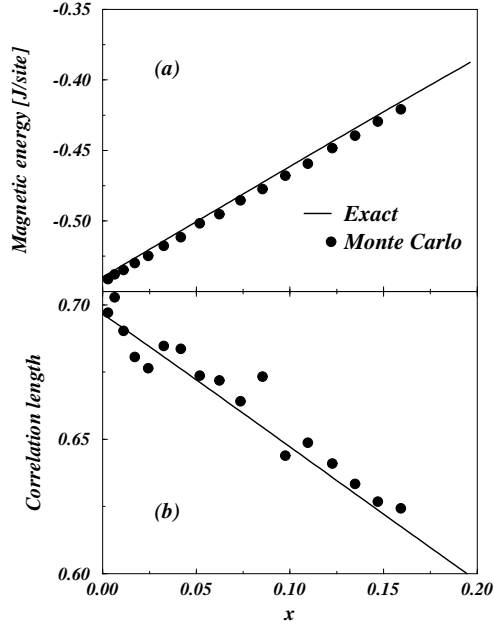}}
\caption{Exact and numerical  Monte Carlo results of magnetic 
energy and correlation length of the spin-spin correlation function,
 vs. hole doping-$x${\protect\cite{thesis}}. Dimer $v$ and $u$, Eq. (\protect{\ref{di}}), 
is used on a  $2 \times 64$ ladder.} 
\label{evmc}  	 
\end{figure}

For dimer doped RVB state, where  
\begin{equation}
 u_{ij}=v_{ij}
=\left\{ \begin{array}{ll}
1  &\;\; |{\bf r}_{ij}|=1 \\
0   &\;\; \mbox{otherwise} 
\end{array} \right. .
\label{di}
\end{equation}
we obtained exact results using transfer matrix technique. For a $4 \times 40$  undoped lattice \cite{mypaper}
the magnetic energy is $E_{mag}=-0.320744\; J/bond$. The Monte Carlo result is 
$-0.3210 \pm 0.0002$. 
$\langle {\bf S}_0 {\bf S}_{r}\rangle$ is exponentially decaying 
with correlation length of  $\xi_{\mbox{dimer}} = 0.724$, the Monte Carlo result is $0.738$.  
Exact and Monte Carlo results for the doped $2\times 64$ ladder\cite{thesis}  appear in Fig.(\ref{evmc}). 
Our program successfully reproduced existing data for RVB states \cite{LDA,japan,thesis}. 
Other tests of the program appear below.

We  also use  the Gutzwiller Approximation (GA)  
to evaluate expectation values in the doped RVB state.
The GA is discussed  in App.\ref{largen}.

\section{The variational parameters}
\label{uuuu}

In the undoped, $x=0$, case we treat three  classes of singlet bond amplitude $u$.
\begin{eqnarray}
u_{p}(r)&=&\frac{1}{r^p} \label{uppl} \\
u_{ex}(r)&=& u_{sr}(r)\frac{1}{r^{0.4}}\exp{\left(-\frac{r}{\xi}\right)}  \label{uex}\\
u_{g}(r)&=& u_{sr}(r)\exp{\left(- Q r^2\right)} \label{ug}
\end{eqnarray}
with $u(1)=1$ and  
\begin{equation}
u_{sr}(r)=a_1\exp{\left(-\frac{r}{\xi_{sr}}\right)} + a_2, 
\label{usr}
\end{equation}
where for $u_{ex}$ ($u_g$) $\xi^{-1}_{sr}=1.7\;(2)$ and  $a_2=0.05\;(0.018)$.  
$u_{sr}$ determines the   short range decay of $u_{ex}$ and $u_g$\cite{thesis}.
We also use $u=u_{MF}$. $u_{MF}$  is derived from the  Schwinger-boson mean field
theory of the Heisenberg model \cite{Book,japan}.
For $x>0$ we use $u_{p}$, Eq. (\ref{uppl}), and $u_{ex}$, Eq. (\ref{uex}).

For the function $v$ the following cases of inter-hole correlations are treated:
\begin{equation}
\begin{array}{llll}
v_{ins}^{\gamma}({\bf r}_{ij})&=& \cases{1& $( i,j) \in \gamma$\cr
0& $(i,j) \notin \gamma$}\\
v^{met}_{\bf r} & =& 1/L^2\sum_{{\bf k}\in \Sigma}  v^{met}_{\bf k} e^{-i{\bf k}\cdot{\bf r}} \\
v_{\alpha}({\bf r}) &=&\sum_{\hat{\eta}} c_\alpha ({\hat{\eta}})  
\delta_{{\bf r}, \hat{\eta}},~~~\alpha=s,d  \\
\end{array}
\label{uv}
\end{equation} 
where $|v^{met}_{\bf k}|=1$, $\hat{\eta}$ are nearest neighbor vectors on the square lattice,
$c_s=1$ and $c_d=\hat{\eta}_x^2-\hat{\eta}_y^2$.

$v_{ins}^{\gamma}$ puts the $N_h$
holes on random sites. This state describes an insulator
with  disordered localized charges.

\begin{figure} 
\centerline{ \epsfysize=3.0cm
\epsfbox{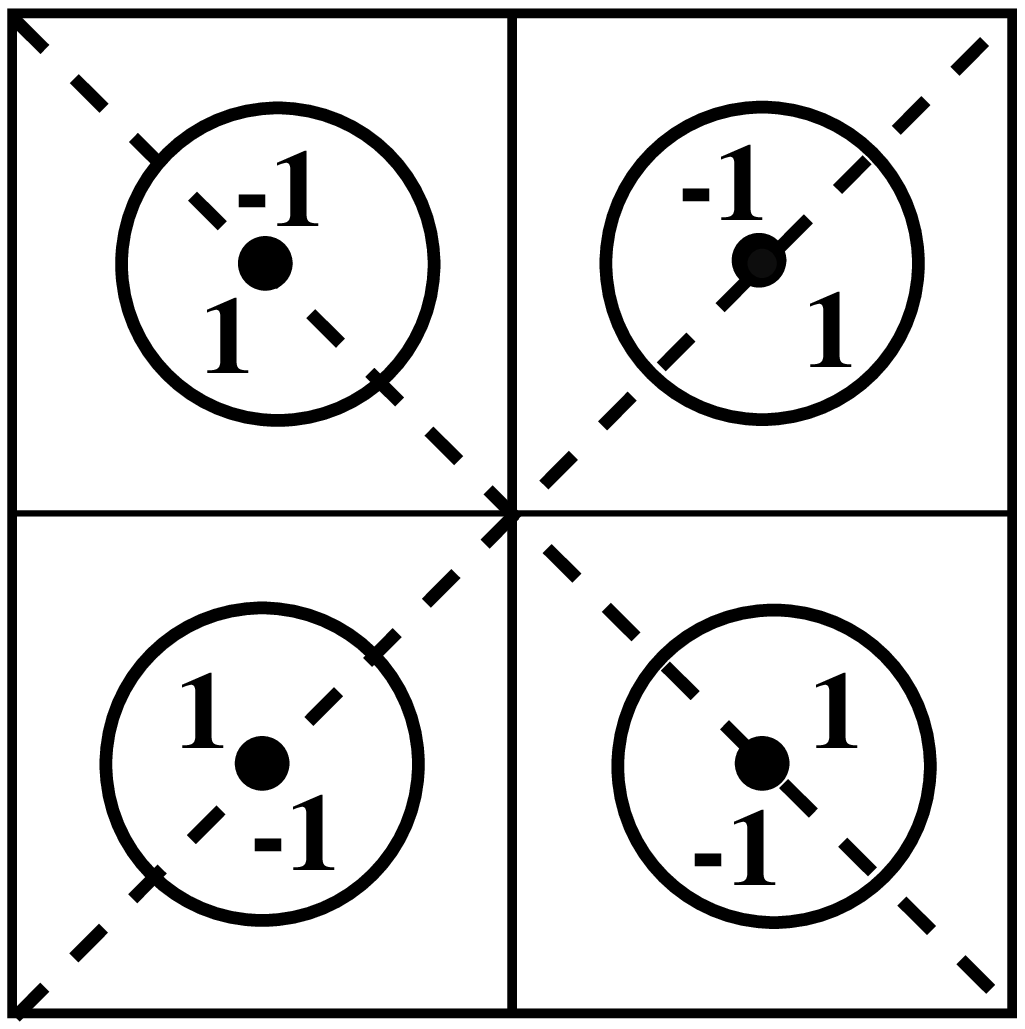}}
\caption{$v=v^{met}_{\bf k}$. $\Sigma$ is centered at Fermi pockets around  
${\bf k}_{min}=(\pm \pi/2,\pm \pi/2)$. 
Within the pockets $v_{met}= \pm1$, otherwise $v_{met}= 0$. 
Note that $v_{met}$ has $d_{x^2-y^2}$ symmetry.}
\label{vsym}  	 	 
\end{figure}

$v_{met}$ describes  weakly interacting holes  in a ``metallic'' state:
\begin{eqnarray}
\prod_{{\bf k}\in \Sigma}
f^{\dagger}_{\bf k}|0\rangle =
{\cal P}_G(x) \exp{\left(\sum_{{\bf k}}v^{met}_{\bf k}
 f^{\dagger}_{\bf k} f^{\dagger}_{-{\bf k}+\left(\pi,\pi\right)}\right)} |0\rangle  
= {\cal P}_G(x)\exp{\left(\sum_{ij}
v_{ij}^{met} f^{\dagger}_i f^{\dagger}_j\right)} |0\rangle 
\label{fgs}  
\end{eqnarray} 
where  the product is over $N_h$ states, 
\begin{eqnarray}
v^{met}_{\bf k} =  \left\{ 
\begin{array}{ll} 
sign({\bf k})  &  {\bf k} \in  \Sigma \\
 0     &   {\bf k} \not\in  \Sigma 
\end{array}
\right. 
\label{vgh}
\end{eqnarray}
and   $v^{met}_{ij}=
\sum_{{\bf k}} v^{met}_{\bf k} e^{-i{\bf k}\left( {\bf r}_ i - {\bf r}_j\right)}$.
Here we  check $\Sigma$ which is centered at  
${\bf k}_{min} = \left(\pm\frac{\pi}{2}, \pm\frac{\pi}{2}\right)$. See Fig. \ref{vsym}.
Results for $\Sigma$ centered at  
${\bf k}_{min} =\left(0, \pm \pi\right),\;\left(\pm \pi, 0\right)$ are {\em not} qualitatively different \cite{thesis}.
$v_{met}$  obey
\begin{equation}
v_{{\bf k} + (\pi,\pi)} = -v_{\bf k},
\label{ssuv}
\end{equation}
hence $v^{met}_{ij}$ only connects   $i\in A$ to $j\in B$. 
Correlations in a state with $v=v_{met}$ 
were previously computed by Bonesteel and Wilkins\cite{BW}. 

$v_s$ and $v_d$ describe tightly bound hole pairs in relative $s$ and $d$-wave 
symmetry respectively.

\section{Order parameters}
\label{orpar}
\subsection{Local magnetic moment and long range magnetic correlations}
The local magnetization is
\begin{equation}
m_0^2 = \frac{1}{L^4}\sum_{ij}  \langle{\bf S}_i {\bf S}_j\rangle e^{-i(\pi,\pi)({\bf r}_i-{\bf r}_j)}
\label{sp}
\end{equation}
where e.g.  $S^+=S^x+iS^y \equiv a^\dagger b$. With respect to
Eq. (\ref{ooooo}), $ \langle{\bf S}_i {\bf S}_j\rangle\equiv S(r_{ij})$
is calculated using\cite{LDA}
\begin{equation}
{\bf S}_i {\bf S}_j(\gamma,\Lambda_{\gamma})_1
=\left\{ \begin{array}{ll}
\pm \frac{3}{4}  &\;\;  i \; j  \mbox{ are on the same loop in } \Lambda_{\gamma}  \\
0   &\;\;\; \mbox{otherwise} 
\end{array} \right. 
\label{csisj}
\end{equation}
where the sign is + if $i$ and $j$ are on the same sublattice. To check our program we also used 
an alternative  procedure to calculate magnetic cerrelations. See App.\ref{c222}.

\begin{figure}
\centerline{ \epsfysize=13.0cm
\epsfbox{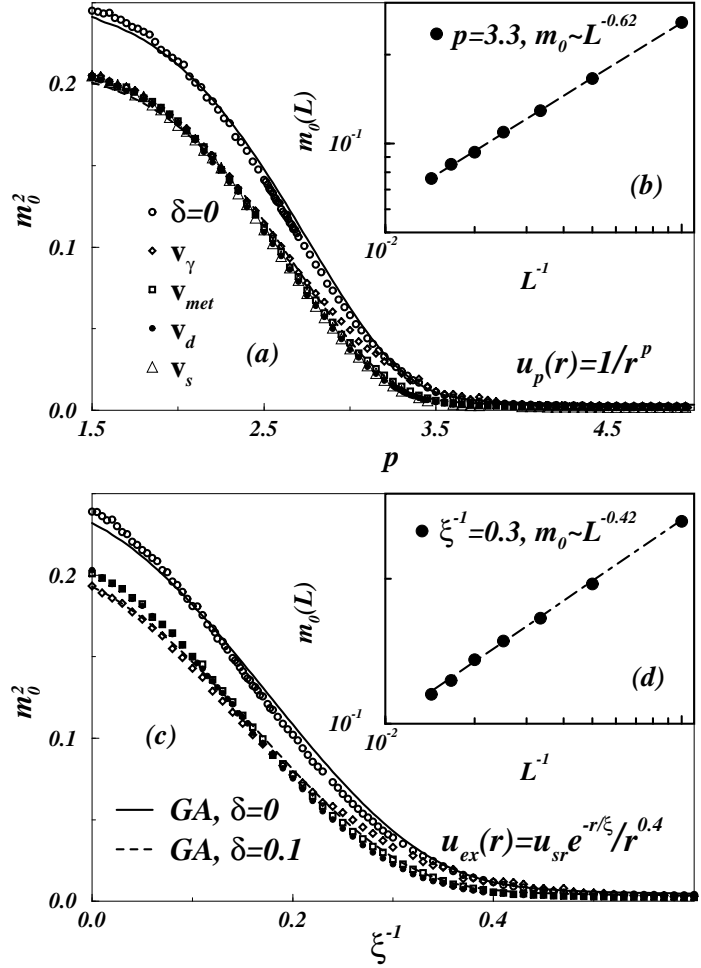}}
\caption{(a) The local magnetization squared, $m_0^2(p)$, of doped and undoped RVB wavefunctions,
versus the variational power $p$, defined by Eq. (\protect{\ref{uppl}}).   
Lattice size is
 $40\times 40$, and in the doped case hole concentration is 10\%.  
Results agree  with
the Gutzwiller approximation (lines).
The hole bond parameters
$v$ are defined in Eq.(\protect{\ref{uv}}).
$m^2_0$ is weakly dependent on $v$.
(b)Finite size scaling of $m_0(L)$ for $p=3.3$ which indicates vanishing local magnetization 
at $L\to \infty$.
(c) $m_0^2(\xi)$ versus the variational correlation length $\xi$, defined by (\protect{\ref{uex}}).
(d)  Finite size scaling of $m_0(L)$ for $\xi^{-1}=0.3$ which indicates vanishing local magnetization 
at $L\to \infty$. }
\label{sppvsp}
\end{figure}

In Fig. \ref{sppvsp}(a)   $m^2_0(p)$ is plotted for $\Psi[u_p;x=0]$ and $\Psi[u_p,v;x=0.1]$
  for various choices of $v$.
Finite size scaling in Fig.\ref{sppvsp}(b) for $x=0.1$ indicates vanishing long range order,
$m_0\to 0$, at $p_c= 3.3$. It lowers the  bound  given previously by Ref. \cite{LDA}:  at $p_c\le  5$.
In Fig. \ref{sppvsp}(c) 
$m^2_0(\xi^{-1})$ is plotted for $\Psi[u_{ex};x=0]$ and $\Psi[u_{ex},v;x=0.1]$.
Finite size scaling in Fig.\ref{sppvsp}(d)  indicates
 $m_0\to 0$, at  $\xi^{-1}= 0.3$.
In all the cases the GA (lines) works well.

\begin{figure}
\centerline{ \epsfysize=8.0cm
\epsfbox{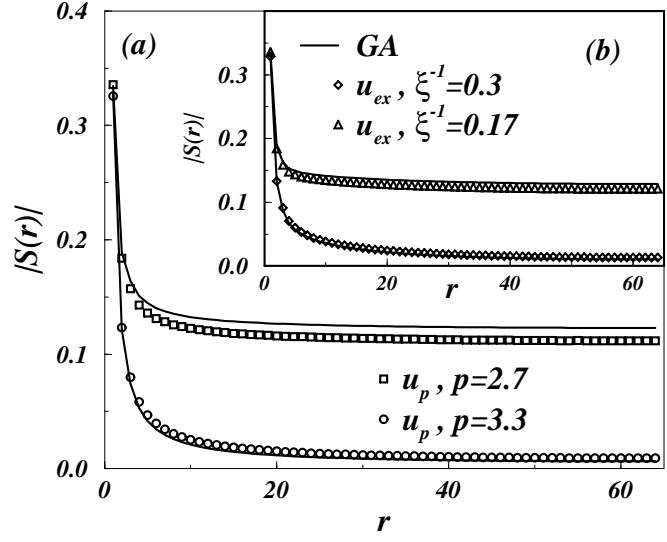}}
\caption{(a) Calculations of spin-spin correlation function ($S(r)$) with Monte Carlo and GA (lines) 
for undoped states with $u=u_p$, Eq. (\protect{\ref{uppl}}). The size of the lattice 
is $L=128$. (b) Calculations of $S(r)$ using $u=u_{ex}$, Eq. (\protect{\ref{uex}}). Note how slow $S(r)$ decays.
In both cases  there is a good agreement between Monte Carlo and GA, 
and $S(r))$ weakly  effected by doping. }
\label{sofr}
\end{figure}

Good agreement between GA and Monte Carlo is also seen in  
Fig.\ref{sofr}(a) and (b), where $S(r)$  is plotted for $u_p$, and 
 $u_{ex}$ respectively. Note how slow $S(r)$ decays for
$\xi^{-1} = 0.3$. By  Fig.\ref{sppvsp}(b) $m_0(L)\sim L^{-0.42}$ in this state.
Exponentially decaying spin correlations are seen, both by Monte Carlo and GA,
for  $u_p$ with $p\gtrsim 3.7$ 
and  $u_{ex}$ with $\xi^{-1}\gtrsim 0.4$  \cite{thesis}.
Details of $u_{sr}$, Eq.(\ref{usr}), have strong effect on long range spin correlations.

We use GA to extrapolate Monte Carlo calculations for $S(r)$.
In App.\ref{exuyy} we find for exponential bond amplitude,
$u(r)= \exp\left(-r / \xi \right)$ and $\xi\gg 1$, that 
$S(r)$ decays exponentially with correlation length
\begin{equation}
\xi_{ex} \propto \exp{\left((1-x ) \frac{3 \pi}{2} \xi^2\right)}.
\label{xiex}
\end{equation}
For Gaussian bond amplitude, $u(r)= \exp\left(- \mu r^2  \right)$ with $ \mu\ll 1$, we find in App.\ref{andel}   that
$S(r)$ decays  exponentially with correlation length
\begin{equation}
\xi_g \propto\frac{1}{\sqrt{\mu}} \exp{\left((1-x )\frac{\pi}{4\mu}\right)}.
\label{xigus}
\end{equation}
For $u_p$, App.\ref{pluan} suggests vanishing long range order,
$m_0\to 0$, at $p_c\le  3$.

Correlation lengths (\ref{xiex}) and (\ref{xigus}) explain the slow decay of $S(r)$ in Fig.\ref{sofr}(b). It also
indicate, that in the $L=\infty$ system, a small change in the ground state parameters brings an extremely sharp change
in long range magnetic correlations.

\subsection{Superconducting order parameter}
\begin{figure}
\centerline{ \epsfysize=7.0cm
\epsfbox{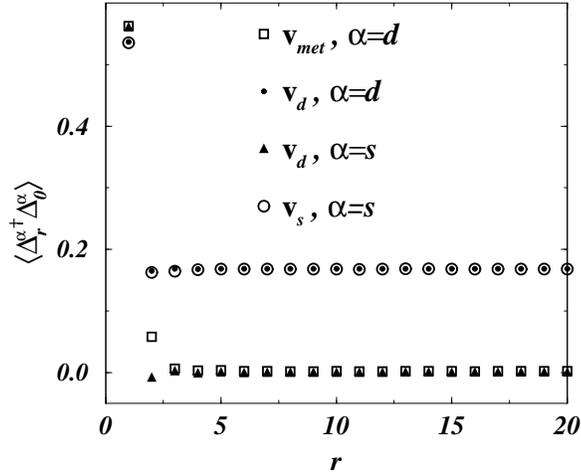}}
\caption{ The singlet pair correlation function. 
$\langle\left(\Delta_{r}^{\alpha}\right)^{\dagger}\Delta_0^{\alpha}\rangle$.
 $\Delta_i^{\alpha}$ defined in Eq. (\protect{\ref{opar}}),  and $\alpha=s,d$. In this graph $u=1/r^{3.3}$.
$\Psi[u,v_{\alpha};x]$ has superconducting (Off-Diagonal) long range order only of symmetry $\alpha$.
$\Psi[u,v_{met};x]$ has no ODLRO in either symmetry. }
\label{odlro}	 
\end{figure}

 The superconducting singlet order parameters are 
\begin{equation}
\Delta^{s,d}_i =\sum_{\hat{\eta}}c_{s,d}(\hat{\eta} ) 
\label{opar}
\end{equation}
where 
\begin{equation}
\Delta_{ij} =  f^\dagger_i f^\dagger_{j} (a_i b_{j}-b_i a_{j})/\sqrt{2}
\label{dalte}
\end{equation}
The expressions of $\Delta$'s matrix elements are discussed in App.\ref{resddd}.
By gauge invariance imposed by the Gutzwiller projector,   $\langle \Delta^{s,d}\rangle=0$. 
However, $\Psi[u,v_{s (d)};x > 0]$ describes 
true $s$ ($d$)-wave superconductors as seen by the singlet pair correlation function 
$\langle\left(\Delta_r^{\alpha}\right)^{\dagger}\Delta_0^{\alpha}\rangle$, $\alpha = s,d$, in Fig.\ref{odlro}.
For $v=v_{\alpha}$, 
$\lim_{r\rightarrow \infty}\langle\left(\Delta_r^{\alpha}\right)^{\dagger}\Delta_0^{\alpha}\rangle\neq 0$ and 
$\Psi$ has (off-diagonal) long range order in $\Delta_{\alpha}$.
In contrast, the insulator states  $\Psi[u,v_{ins},x]$ and the ``metallic'' states 
$\Psi[u,v_{met},x]$ have no long range superconducting order 
of either symmetry.

\section{Effective Hamiltonians}
\label{efh}
\subsection{Magnetic energy and related parameters}
\label{emag}
Magnetic order is driven by the diluted antiferromagnetic quantum Heisenberg model
\begin{equation}
{\cal H}^J =J\sum_{\langle ij\rangle} {\bf S}_i {\bf S}_j.
\label{qhaf}
\end{equation}

\begin{figure}
\centerline{ \epsfysize=14.0cm              
\epsfbox{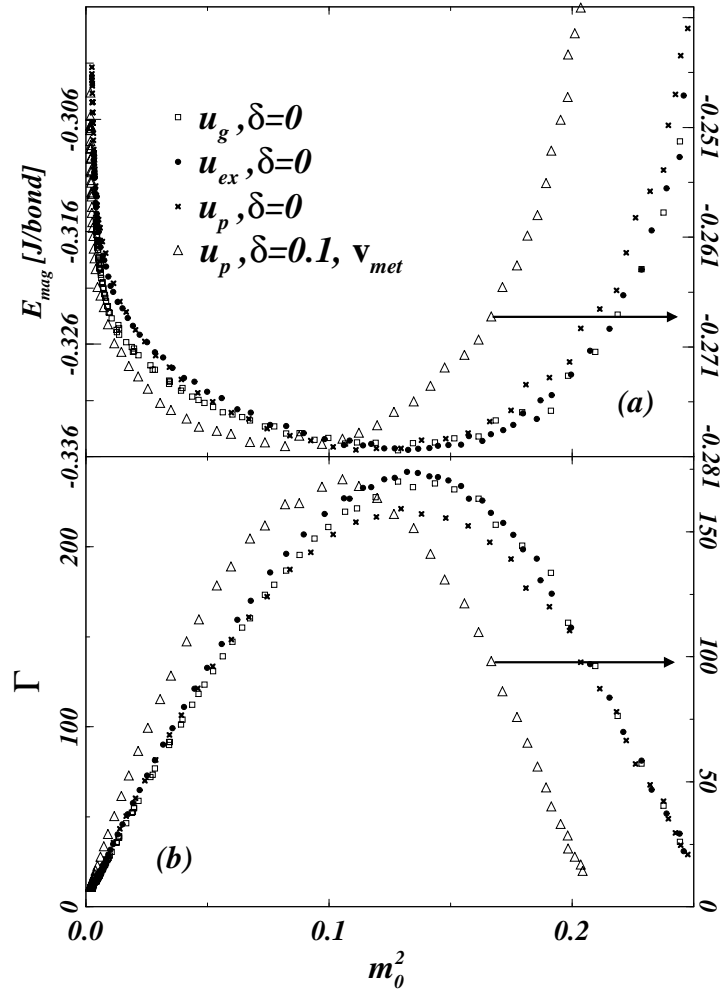}}
\caption{(a) Magnetic energy ($E_{mag}$) 
versus local magnetization squared $m_0^2$. 
Variational parameters are : for $x=0$(left $y$ scale), $u_p(p)$ 
Eq. (\protect{\ref{uppl}}),
$u_{ex}(\xi)$ Eq. (\protect{\ref{uex}}),   and $u_g(Q)$ Eq. (\protect{\ref{ug}}). 
For $x=0.1$ (right $y$ scale) $v_{met}$ and $u_p$. 
lattice size is  $L=40$. 
The optimal parameters  for each $u$ appear in Table \protect{\ref{paraatmin}}.
The minimal magnetic energy for $x=0$ is $-0.335\pm 0.0005 J/bond$. 
(b) The average density of a loop per site $\Gamma$, Eq. (\protect{\ref{gbar}}), versus $m_0^2$ for variational cases as in (a).   
In all the cases $\Gamma$ is correlated with $E_{mag}$.}
\label{gammas}
\end{figure}

{\em Magnetic energy for $x=0$}: 
In Fig. \ref{gammas}(a) $E_{mag}(p)$, $E_{mag}(\xi)$ and  $E_{mag}(Q)$  
are plotted as a function of $m_0^2(p)$, $m_0^2(\xi)$
and $m_0^2(Q)$ for $u_{ex}$, $u_p$ and $u_g$ , Eqs. (\ref{uppl}), (\ref{uex}) and (\ref{ug}) respectively.
In $x=0$, {\em all} the three bond amplitude yield lowest magnetic energy of
\begin{equation}
E_{0} = -0.335 \pm 0.0005 J/bond,
\label{emin}
\end{equation}
For $u_p$, the optimal value of $p$ is $p_{optimal}=2.7$, and $m_0^2(p=2.7)=0.105\pm0.005$.
The ground state parameters of the Heisenberg model on an $L=40$ lattice are : E(ground state)= 0.3347 J/bond 
and $m_0^2$(ground state)=0.109\cite{Sandvik}.
Table \ref{paraatmin} contains a summary of 
results for the optimal choice of parameters in all the classes.

\begin{figure}
\centerline{ \epsfysize=8.0cm
\epsfbox{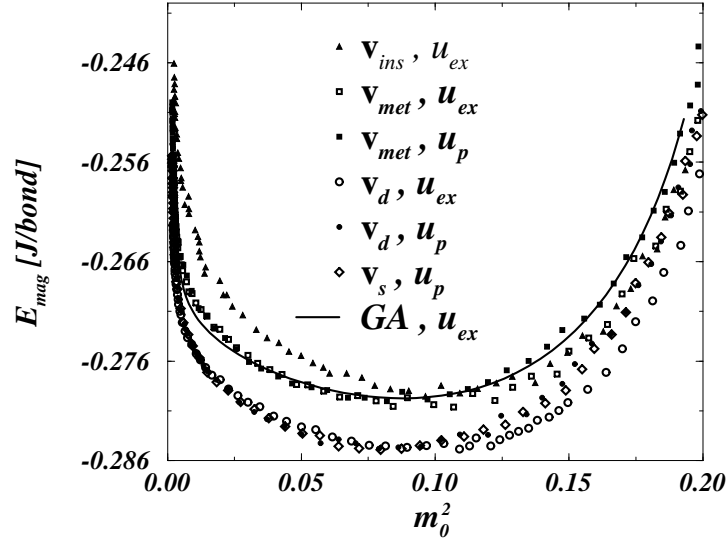}}
\caption{Magnetic  energy  ($E_{mag}$),  
for  $u_p$  Eq. (\protect{\ref{uppl}}), and   $u_{ex}$  Eq. (\protect{\ref{uex}}),
  versus local magnetization squared $m_0^2$,  using various hole distributions from Eq. (\protect{\ref{uv}}).
 The  density of holes is $x=0.1$ and lattice size is  $L=40$.
$E_{mag}$ is weakly dependent on inter-hole correlations.
For $u_{p} \;(u_{ex})$, $E_{mag}$ is minimized at $m_0^2(p) \approx 0.08\;(0.1), p=2.7\;(\xi^{-1}=0.17)$.}
\label{hemag}  	 	 
\end{figure}

{\em Magnetic energy   for $x=0.1$}: 
In Fig.\ref{hemag}   $E_{mag}(p)$ and $E_{mag}(\xi)$  are plotted as a function of   $m_0^2(p)$ and $m_0^2(\xi)$,
for $x=0.1$  and various choices of $v$ from Eq. (\ref{uv}).
Within numerical errors,
 all states minimize ${\cal H}^{J}$ at the same optimal parameters as for $x=0$ (Table \ref{paraatmin}).
For  $u_p$, by Fig. \ref{sppvsp}(a) it yields local magnetization 
of $m^2_p(x=0.1)= 0.08$. 
For $u_{ex}$, Fig. \ref{sppvsp}(c)  shows 
$m^2(x=0.1)= 0.1$.  
Thus we conclude that aside from the trivial kinematical
constraints, {\em the hole density and  correlations have 
little effect on the magnetic energy at low doping.}

 A better understanding of the properties of the optimal 
bond amplitude for ${\cal H}^J$ is gained by  the {\em average loop density} defined below.
From Eq. (\ref{csisj}), a DLC contributes to $m_0^2$, Eq. (\ref{sp}), it's number of pairs of sites,
which share the same loop hence
\begin{equation}
m_0^2 = \frac{3}{4L^4}
 \overline{\left(\sum_{\lambda \in \Lambda_{\gamma}}l_{\lambda}^2\right)}=\frac{3}{4L^4}
 \overline{\left(\sum_{i\not\in\gamma}l_{\lambda_i}\right)} 
\label{allps}
\end{equation}
where  $l_{\lambda}=\sum_{i \in \lambda}1$ is the loop length, and $i \in \lambda_i$.
Thus $L^2 m_0^2=S(\pi,\pi)$ is proportional to the average loop length per site.

The average radius of gyration of a loop is:
\begin{equation}
r_g \equiv \overline{\left(\frac{1}{n_{\Lambda}}\sum_{\lambda \in \Lambda_{\gamma}} r_{g}^{\lambda}\right)},
\label{rgg}
\end{equation}
where
\begin{equation}
\left(r_{g}^{\lambda}\right)^2 =
\frac{1}{l_{\lambda}} \sum_{i\in \lambda}\left({\bf r}_i - {\bf r}_{cm}^{\lambda}\right)^2=
\frac{1}{2l_{\lambda}^2} \sum_{i,j\in \lambda}\left({\bf r}_i - {\bf r}_j\right)^2,
\label{rgl}
\end{equation}
with $r_{cm}^{\lambda}=\frac{1}{l_{\lambda}}\sum_{i\in \lambda} {\bf r}_i$, 
and $n_{\Lambda}$ is the number of loops in the DLC $\Lambda$.
With  Eqs. (\ref{allps}) and (\ref{rgg}) we  define the average density of a loop per site
\begin{equation}
\Gamma \equiv \frac{L^2 m_0^2}{r_g^2}.
\label{gbar}
\end{equation}

The average loop density, $\Gamma$,
is plotted in Fig.\ref{gammas}(b),
in the undoped case for all the bond amplitudes  (\ref{uppl}), (\ref{uex}), and (\ref{ug}); and in the doped case for
$\Psi[u_p, v_{met};x=0.1]$. 
Comparison with Fig.\ref{gammas}(a) shows that $\Gamma$ is correlated with the magnetic energy.
For vanishing $m_0$, $\Gamma$ converge to its value 
in  the dimer RVB state, Eq. (\ref{di}), where $\Gamma(\mbox{dimer RVB})\approx 9.6$. This value of $\Gamma$
is only slightly larger than
$\Gamma$'s value for an ensemble of DLCs, which include only configurations with two (or four) sites loops with dimer bonds.
For such loops $r^{\lambda}_g=0.5$ (or $\sqrt{2}/2$) and $\Gamma=l_{\lambda}/ r_{g,\lambda}^2=8$.

\begin{figure}
\centerline{ \epsfysize=7.0cm
\epsfbox{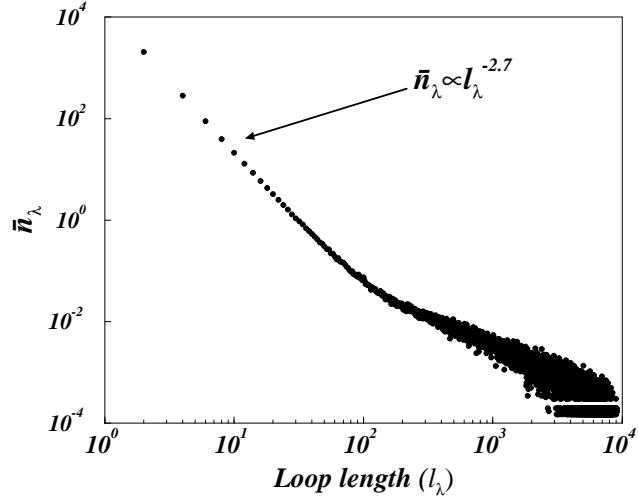}}
\caption{The average number of loops ($\bar{n}_{\lambda}$), versus the number of sites on a loop ($l_{\lambda}$).
  The state is with $u=u_{MF}$, which is derive from the Schwinger bosons mean field theory of ${\cal H}^J$. 
The size of the lattice is $L=128$. For $l_{\lambda} {\protect\lesssim} 130$,  $n_{l} \propto l^{-2.7}_{\lambda}$. }
\label{histogram}
\end{figure}

The occurrence of loop lengths ($l_{\lambda}$) is interesting. In Fig.\ref{histogram} we plot an histogram
of the number of loops ($\bar{n}_{\lambda}$), versus the number of sites on a loop ($l_{\lambda}$).
 The size of the lattice is $L=128$, and
$u=u_{MF}$, which is derive from the Schwinger bosons mean field theory of ${\cal H}^J$ \cite{Book}. 
 For  all the bond amplitudes and lattice sizes we have checked $\bar{n}_{\lambda}(l)$ 
 decays either algebraically or exponentially.

\subsection{Single hole hopping energy}

A single hole hopping in the antiferromagnetic 
background has been shown by semiclassical
arguments\cite{t-JSC,Book}, to be effectively restricted at low energies
to hopping between sites on the same sublattice:
\begin{equation}
{\cal H}^{t'}= \sum_{\langle i k\rangle\in A,B}t_{ik}' f^\dagger_i f_k (a^\dagger_k a_i +b^\dagger_k b_i) 
\label{tprime}
\end{equation}
where $i,k$ are removed by two adjacent lattice steps, and $t' > 0$.
Unconstrained, the single hole ground state of $H^{t'}$ has momentum
on the edge of the magnetic Brillouin zone, in agreement with
exact diagonalization of $t-J$ clusters\cite{dagotto}.
Previous investigations have found that {\it inter}-sublattice hopping
(the $t$-term in the $t-J$ model), is a
high energy processes in  the AFM correlated state\cite{t-JSC,Book}. We thus expect
the same to hold even in 
RVB spin liquids with strong short range AFM correlations but
no long range order. The primary effects at low doping may be to shift
the ordering wavevector.

We denote by $t'_d$ ($t'_h$) the coefficients of second (third) nearest neighbor  hopping terms. 
For $t'_h>t'_d/2$ the  single hole bend minimum
is at ${\bf k}_{min}=(\pm\pi/2,\pm\pi/2)$, otherwise
${\bf k}_{min}=\pm(0,\pi),\pm(\pi,0)$.  Here we put
$t'_h=1$, $t'_d=0.5$.

\begin{figure}
\centerline{ \epsfysize=7.0cm
\epsfbox{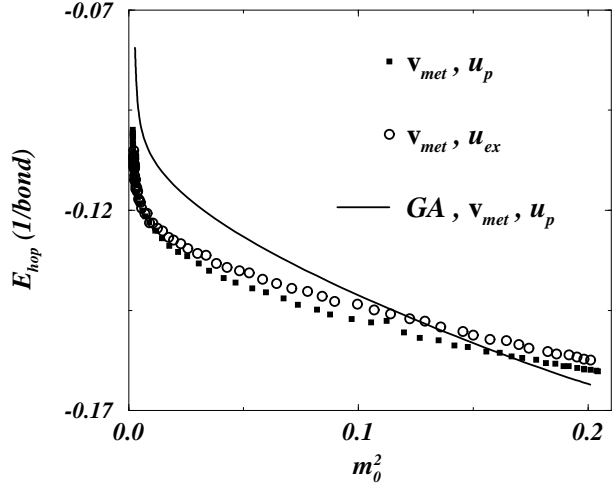}}
\caption{The  single hole hoping energy  $E_{hop}$, Eq. (\protect{\ref{tprime}}), 
 versus local magnetization squared $m_0^2$ for metalic hole distributions, 
and spin bond amplitudes $u_p$ and $u_{ex}$.
The  density of holes is $x=0.1$ and lattice size is  $L=40$.
In $H^{t'}$, $t'_h=1$, $t'_d=0.5$, and the single hole band's minimum is at 
 ${\bf k}_{min} = (\pm \pi/2,\pm \pi/2)$). 
The single hole hopping 
prefers  longer range $u(r)$ and hence higher local magnetic moment. 
It also pefer metallic states over $v=v_s, v_d$ {\protect\cite{thesis}}.}
\label{ehop}  	 	 
\end{figure}

Results for the expectation value of  $H^{t'}$ are plotted in Fig.\ref{ehop}. 
The single holes hopping, Eq. (\ref{tprime}), 
prefers the metallic states $v=v_{met}$  over states with
$v=v_{s}, v_d$\cite{thesis}.  It also
prefers longer range $u(r)$ and thus actually {\em enhances} magnetic order
at low doping. This is a type of
a Nagaoka effect, where mobile holes separately polarize 
each of the  sublattices ferromagnetically.

\subsection{The double hopping energy}
\label{douh}

\begin{figure}
\centerline{\epsfysize=2.0cm
\epsfbox{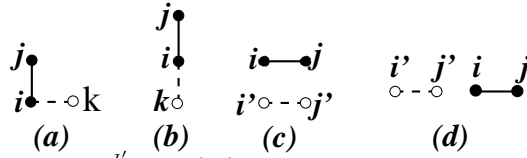}}
\caption{(a), (b), and (c) are the terms of $H^{J'}$, Eq. (\protect{\ref{jprime}}). 
Dashed line and empty circles = $\Delta$. 
Term (a) is a rotation of a singlet pair, it distinguishes between
 $s$ to $d$ wave  superconducting order parameters. Term (c) prefers $v_{d(s)}$ over metallic states with 
$v_{met}$. 
Term (d) dependency on the variational parameters is similar to that of  (c), it is excluded due to thermal noise.}
\label{duforms}	 
\end{figure}

 We consider Cooper pairs hopping terms
\begin{equation}
{\cal H}^{J'}=-J'\left(\sum_{ijk}\Delta^{\dagger}_{ij}\Delta_{ik}+
\sum_{\langle i j\rangle, i'j'}\Delta^{\dagger}_{ij}\Delta_{i'j'}\right)
\label{jprime}
\end{equation}
Calculation of ${\cal H}^{J'}$ matrix elements is discussed in App.\ref{resddd}.
The first term in ${\cal H}^{J'}$ is derived from the large $U$ Hubbard model to order
$J' =t^2/U$\cite{Book}. It includes terms (a) and (b) in Fig.\ref{duforms}. Term (a) is 
a {\em rotation} of the singlet pair. 
 It is positive for $v_s$ \cite{thesis} and hence 
prefers $v_d$ over $v_s$. Term (c) in Fig.\ref{duforms} is a  parallel {\em translation}
of singlets. It  prefers superconductivity
with $v=v_d$ or $v=v_s$ over metallic states with $v=v_{met}$  \cite{comm-PH}. 
For $x=0.1$, ${\cal H}^{J'}$ is minimize by $v_d$.

\begin{figure}
\centerline{\epsfysize=7.0cm
\epsfbox{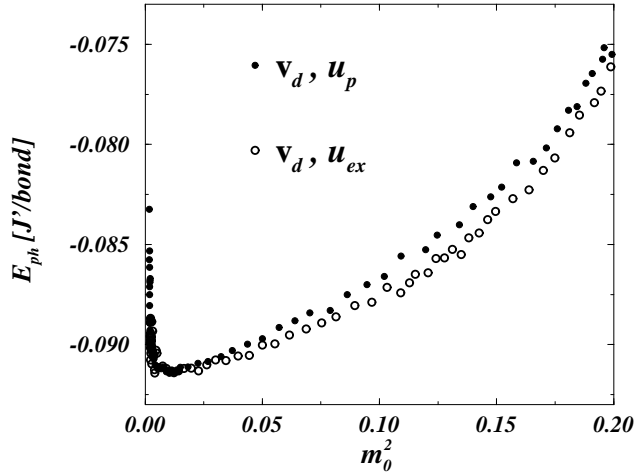}}
\caption{The expectation value of $H^{J'}$, Eq. (\protect{\ref{jprime}}), $E_{ph}$, 
versus $m_0^2$, for $u_p$ and $u_{ex}$. 
In contrast to the magnetic energy Fig.(\protect{\ref{hemag}}), 
$E_{ph}$ prefers a {\em vanishing} $m_0$ at $L \to \infty $. 
Note how similar the graphs are for $u_{ex}$ and $u_p$.}
\label{eph}	 
\end{figure}

In Fig. \ref{eph}    the ground state energy $E_{ph}$ of (\ref{jprime}) is
plotted for $v=v_{d}$, $u=u_p$ and $u=u_{ex}$. $x$=0.1,  and the size of the lattice is $L$=40.
The variational energy is minimize
at $p=3.35$ and $\xi^{-1}=0.35$, for $u_{ex}$ and $u_p$, respectively.
In both cases, by the finite size scaling of Fig.\ref{sppvsp}(b) and (d), it
indicates vanishing $m_0$ at  $L\rightarrow \infty$. 
{\em Thus, Cooper pair hopping drives the groundstate toward a spin liquid phase!}

The Gutzwiller approximation  fails to predict this effect.  
According to the GA, the minimum of the double hopping energy roughly coincides with
the minimum of the magnetic energy ($\langle{\cal H}^{J}\rangle$). This is understood by (see App. \ref{largen}):
\begin{eqnarray*}
\langle S^+_i S^-_j \rangle_{GA}
= - \langle a^{\dagger}_i b^{\dagger}_j \rangle^2_{GA},
\end{eqnarray*}
where $i\in A$, $j\in B$. The GA agrees with Monte Carlo results for 
matrix elements of {\em long range} pair hopping.

The matrix element of (d) in Fig.\ref{duforms}, and $\langle n_i^f n_j^f \rangle$ also
drives the groundstate toward a spin liquid, and prefer superconducting over metallic
 states\cite{thesis}. These terms are excluded due to relatively large thermal noise.

\section{A relation between superconducting T$_c$ and local magnetization}
\label{tcm0}
Since ${\cal H}^{J'}$ is the effective model which
drives superconductivity it produces phase stiffness,
which in
the continuum approximation is given by
\begin{equation}
{\cal H}^{J'}\approx {{V_0}\over{2}} \int  d^2 x (\nabla\phi_i)^2
\label{hphi}
\end{equation}
The stiffness constant $V_0$ can be determined variationally from the doped
RVB states.
Imposing a uniform gauge field twist on $\Delta$,
$\Delta_{i,j}\to \Delta_{i,j} exp(i(x_i+x_j) \phi/2L)$, $\langle{\cal H}^{J'}\rangle$  becomes, to second order in $\phi/L$,
\begin{eqnarray}
E_{ph}&=&\frac{V_0\phi^2}{2} \nonumber \\
V_0&=&\frac{d^2 E_{ph}}{d\phi^2}=2J'\left(\langle\Delta^{\dagger}_{0,\hat{y}}\Delta_{0,\hat{x}}\rangle+
\langle\Delta^{\dagger}_{0,\hat{x}}\Delta_{0,-\hat{x}}\rangle
+\langle\Delta^{\dagger}_{0,\hat{y}}\Delta_{\hat{x}+\hat{y},\hat{x}}\rangle\right)
\end{eqnarray}
Following  Ref. \cite{emkivsfs}, at low doping for the square lattice  $V_0$ is roughly 
equal to $T_c$. 

\begin{figure}
\centerline{ \epsfysize=7.0cm
\epsfbox{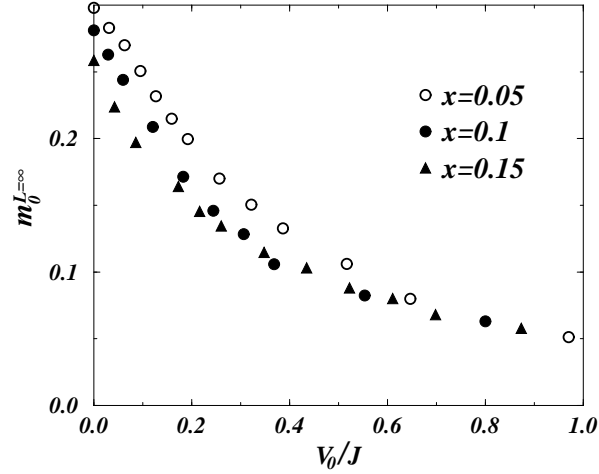}}
\caption{The relation between
thermodynamic local magnetization $m^{L=\infty}_0$ and superconducting  
phase stiffness $V_0$
(related to $T_c$, see text). $u=u_p$, Eq. (\protect{\ref{uppl}}).
$J$ is the Heisenberg exchange energy.  The points are considered 
{\em upper bounds} on $m_0$, which, for $u_p$,  may even vanish for $V_0/J \ge 0.2$.
For $u_{ex}$, Eq. (\protect{\ref{uex}}), $m_0$ vanishes for $V_0/J \ge 0.5$.}
\label{jproverj}  	 	 
\end{figure}

In Fig.(\ref{jproverj}) we show our main result: The staggered magnetization $m_0(p)$ 
for  ${\cal H}^{J}+{\cal H}^{J'}$ is plotted against the 
superconducting to magnetic stiffness ratio  $V_0(p)/J$ for different doping concentrations $x=0.05,0.1,0.15$,
 $v=v_d$, and $u=u_p$. 
The actual free parameter in the graph is $J'/J$, from which $m_0$ and $V_0$ are determined 
variationally.
Two primary observations are made: (i) The local magnetization is 
sharply reduced at relatively low superconducting stiffness  (and  $T_c/J$). 
(ii) The relation between $m_0$ and $V_0/J$ appears to be  independent of $x$.

For $u_{ex}$, Eq. (\ref{uex}), it requires $V_0(\xi)/J=0.49$ for   $\langle {\cal H}^{J}+{\cal H}^{J'}\rangle$ to be
 minimized at $\xi^{-1}=0.3$. By Fig.\ref{sppvsp}(b)  this leads to  $m_0^{L=\infty}=0$.

\section{Summary and Discussion}
\label{diss}

In this paper we used extensive Monte Carlo calculations to study properties of hole doped RVB states.
We found that an effective  model which include Heisenberg and pair 
hopping terms is consistent with  the experimental connection between 
superconductivity and  reduction of local magnetic moment. Within checked variational options    
we showed that the properties of the model
are independent of particular  choice of parameters for the state.
Gutzwiller mean field approximation for magnetic correlations was found to agree with Monte Carlo calculation, 
and used for analytical   extrapolation of  numerical results.
We showed that long range magnetic correlations in RVB states are extremely sensitive to
variational parameters.  We found that the  average loop density is well correlated with the magnetic energy.
We conclude this paper in several  arguments and insights regarding our results.

{\em Magnetic energy and long range magnetic correlation:} 
Note the contrast between correlation lengths (\ref{xiex}) and (\ref{xigus}), and the ``shallowness'' of the minima of
the magnetic energy in Fig.\ref{hemag}. It imply that 
 a very weak pair hopping term in the Hamiltonian 
 causes a dramatic change in long range magnetic correlations.

\begin{figure}
\centerline{ \epsfysize=3.4cm
\epsfbox{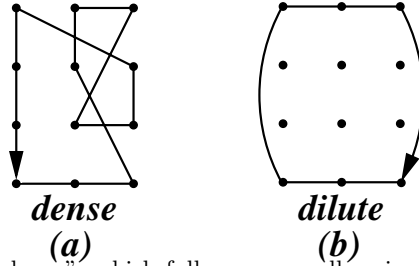}}
\caption{Two kind of loops:
(a) ``Dense loops'', which fully cover small regions of the lattice, and many nearest neighbor pairs.
Bond amplitudes $u$ which maximize the weight ($\Omega_{\Lambda}$) 
of loop configurations with such loops minimize the magnetic energy.
(b) ``Dilute loops'', which  contributes very few nearest neighbor bonds to the magnetic energy, 
Eq. (\protect{\ref{csisj}}).
Loop (a) is denser, in the sense that  it covers more sites  on roughly the same  
``area''$\equiv\left(r_{g}^{\lambda}\right)^2$, Eq. (\protect{\ref{rgl}}).}
\label{tkol}
\end{figure}

{\em Magnetic energy and loop density:}   A comparison between loops (a) and (b) in
Fig.\ref{tkol} shows that large amplitude ($\Omega_{\Lambda}$)
of  DLCs with ``denser'' loops enhance the probability to find 
nearest neighbor sites on the same loop and reduce the magnetic energy. 

The loop density shows that the optimal bond amplitude is determined by an intricate balance 
between $m_0$ and $r_g$. This  relates  {\em quantum spin fluctuations} 
to the average loop density of the ensemble.

{\em Effective model for doped system:} 
${\cal H}^J+ {\cal H}^{J'}$ describes the low energy physics of the lightly doped Cuprates.
As the lattice is doped, its variational ground state is a d-wave superconductor, with a sharply reduced
local magnetic moment.  The model includes built-in pairing. Such a model is supported by the existence of a pseudogap in the 
normal state of the high T$_{\mbox{c}}$ materials.

{\em Relation between phase stiffness and local magnetization:}
Because of finite size uncertainty, $m^{L=\infty}_0$ in Fig.(\ref{jproverj})  is
an {\em upper} bound on the thermodynamic local magnetization. 
A sharper reduction of the local magnetization occurs if:\newline
(a) The GA result of App.\ref{pluan}, $m_0(u_r=r^{-3})=0$, 
is correct to the discrete lattice. In that case $m_0$  vanishs
already at $V_0/J \ge 0.2$.\newline
(b) In finite doping the optimal bond amplitude for ${\cal H}^J+ {\cal H}^{J'}$  decays exponentially. 
In that case $m_0$  vanishes for $V_0/J \gtrsim 0.5$. Variationally, we can not rule out this possibility.\newline
In both of these cases there is a
qualitative agreement with the doping dependence of the local magnetization
and T$_c$,  as measured  by Refs. \cite{Wakimoto,Niedermayer}.

Useful conversations with  C. Henley, S. Kivelson
and  S-C. Zhang, are gratefully acknowledged.  
MH thanks Taub computing center for support.  AA is supported by the Israel Science
Foundation  and the Fund for Promotion of Research at Technion.

\appendix

\section{ The Fermion part of  the dopped RVB state}
\label{ferpart}

The fermion part of $\mid \! \Psi[u, v,x]\rangle$ is
\begin{eqnarray}
\mid \! \Psi(x) \rangle_f = 
P_G(N_h) \exp{\left[\sum_{i\in A,j\in B}
v_{ij} f^{\dagger}_i f^{\dagger}_j\right]} |0\rangle
= \frac{1}{\frac{{N_h}}{2} !} \left[\sum_{i\in A,j\in B}
v_{ij} f^{\dagger}_i f^{\dagger}_j\right]^{\frac{{N_h}}{2}} |0\rangle 
\label{fs1}
\end{eqnarray}
where $N_h=x L^2$. We  write this state as
\begin{equation}
 |\Psi (x)\rangle_f
\equiv \sum_{\gamma} 
C(\gamma)\prod_{k=1}^{N_h/2}
f^{\dagger}_{i_k} f^{\dagger}_{j_k} |0\rangle
\label{fsd}
\end{equation}
where $\gamma$ is a {\em distinct} configurations of $N_h$ holes sites:
\begin{eqnarray*}
\gamma \equiv \left\{\left(i_k\in A, j_k\in B\right)\right\}_{k=1}^{\frac{N_h}{2}}:
 \forall k < k' \;\; i_k < i_{k'},\;  j_k <  j_{k'}.
\end{eqnarray*}
From Eq.(\ref{fs1})
\begin{eqnarray*}
|\Psi(N_h) \rangle_f =
\frac{1}{\frac{N_h}{2} !} \sum_{ \gamma }
v_{{\alpha}_1 {\beta}_1} \cdot \; \cdots \; \cdot  v_{{\alpha}_{\frac{N_h}{2}} {\beta}_{\frac{N_h}{2}}} 
f^\dagger_{{\alpha}_1} f^{\dagger}_{{\beta}_1}
\cdot \; \cdots \; \cdot  f^{\dagger}_{{\alpha}_{\frac{N_h}{2}}}
f^{\dagger}_{{\beta}_{\frac{N_h}{2}}} |0\rangle
\end{eqnarray*}
where  $ \alpha_l \in \left\{i_k\right\}_{k=1}^{{N_h}/2}, \; \beta_l \in \left\{j_k\right\}_{k=1}^{N_h/2}$.
For each $\gamma$ we commute {\em pairs} of operators, without any affect of sign, and
order $A$ holes operators in an increasing order of their site index
\begin{eqnarray*}
|\Psi(N_h) \rangle_f = \sum_{ \gamma }
\sum_{\sigma}
v_{i_1 j_{\sigma(1)}}  \cdot \; \cdots \; \cdot  v_{i_{\frac{N_h}{2}}  j_{\sigma({\frac{{N_h}}{2}})}}
f^{\dagger}_{i_1}f^{\dagger}_{j_{\sigma(1)}}
\cdot \; \cdots \; 
\cdot  f^{\dagger}_{i _{\frac{{N_h}}{2}}}f^{\dagger}_{j_{\sigma({\frac{{N_h}}{2}})}} \mid \! 0\rangle
\end{eqnarray*}
where $\sigma = \sigma\left(N_h/2 \rightarrow N_h/2 \right)$.
Commuting the $B$ operators we get
\begin{eqnarray*}
C(\gamma) =\det\! V\left(\gamma\right)
\end{eqnarray*}
where $V$ is an $N_h/2\times N_h/2$ matrix with 
\begin{eqnarray*}
V\left(\gamma\right)_{kl} \equiv v_{i_k j_l}.
\end{eqnarray*}
When $v$ connects same-sublattice sites the determinant is replaced by a Pffian\cite{thesis}.

\section{Expressions for expectation values}  
\label{esr}
\subsection{Alternative  calculation of  magnetic correlations}   
\label{c222}  
We use the operator identity \cite{Book}
\begin{eqnarray}
-\Delta^{\dagger}_{ij}\Delta_{ij} =
 \left[{\bf S}_i {\bf S}_j - \frac{1}{4}\right](1-f^{\dagger}_i f_i)(1-f^{\dagger}_j f_j),
\label{ida}
\end{eqnarray}
where $\Delta_{ij}=f^{\dagger}_i f^{\dagger}_j (a_i b_j - b_ia_j)/\sqrt{2}$,
to express $\langle {\bf S}_i {\bf S}_j \rangle$.

show that for $i\in A, j\in B$ and $u(r)>0\;\forall r$ 
\begin{equation}
{\bf S}_i {\bf S}_j\left(\gamma,\Lambda_{\gamma}\right)_2 -\frac{1}{4}+\frac{x}{2}
=\left\{ \begin{array}{lll}
-\frac{1}{2}
\left(\sum\limits_{
\scriptstyle (l,n)\in \alpha \atop 
\scriptstyle l \neq i}
\frac{u_{lj}u_{in}}{u_{ln}u_{ij}}
+2\right)  & \;\;\;   (i,\;j) \mbox{ is a bond in } \alpha\left(\Lambda_{\gamma}\right)  \\
\frac{1}{4} & \;\;\;   i,j \in \gamma \\
0  & \;\;\; \mbox{otherwise.} 
\end{array} \right. 
\label{csisj222}
\end{equation}
$\alpha\left(\Lambda_{\gamma}\right)$ is the set of forward bonds  in 
$\Lambda_{\gamma}$, of the sites on sub-lattice $A$.

We demonstrate  Eq. (\ref{csisj222}) for half filled  lattice ($\langle f^{\dagger}_i f_i\rangle=0$).  
With $M_{ij} \equiv (a_i b_j - b_ia_j)$,
\begin{equation}
 M_{ij}M^{\dagger}_{ij}|0\rangle = 2 \; |0\rangle
\label{z1}
\end{equation}
\begin{equation}
M_{ij} M^{\dagger}_{ik}M^{\dagger}_{mj} |0\rangle= M^{\dagger}_{mk}|0\rangle,
\label{z2}
\end{equation}
and hence
\begin{eqnarray*}
M^{\dagger}_{ij} M_{ij}|\Psi[u]\rangle = 
\sum_{\alpha_{ij}}
 \left[\left(\sum\limits_{
\scriptstyle (l,n)\in \alpha_{ij} \atop 
\scriptstyle l\neq i}
\frac{u_{lj}u_{in}}{u_{ln}u_{ij}}\right)
+2\right] \left(\prod_{(l,n)\in \alpha_{ij}}u_{ln}\right)
|\alpha_{ij}\rangle 
\end{eqnarray*}
where $|\alpha_{ij}\rangle$ is a valence bond state, with $(i,j)\in\alpha_{ij}$.

\begin{figure}
\centerline{\epsfysize=5.5cm
\epsfbox{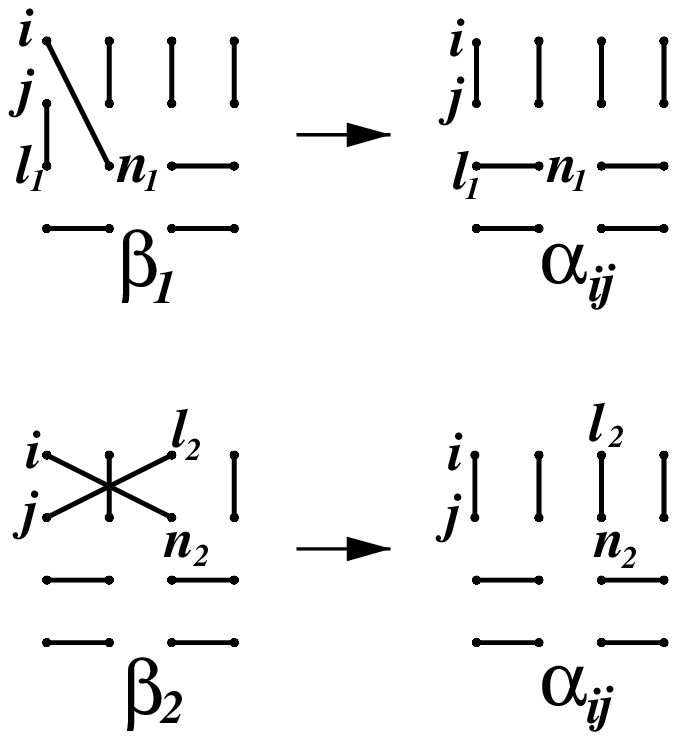}}
\caption{ According to Eq. (\protect{\ref{z2}}), 
$M^{\dagger}_{ij}M_{ij}|\beta_1\rangle=M^{\dagger}_{ij}M_{ij}|\beta_2\rangle=|\alpha_{ij}\rangle$. 
$M_{ij}\equiv a_ib_j-b_ia_j$.}
\label{tconf}	 
\end{figure}

The term in the square brackets  requires further explanation. 
From  Eq. (\ref{z2}), 
for any pair $(l,n) \in \alpha_{ij}:\; l \neq i$, 
$|\alpha_{ij}\rangle=M^{\dagger}_{ij}M_{ij} |\beta\rangle$, where $(i,j),\;(l,n)\not\in\beta$, 
$(l,j),\;(i,n)\in \beta$, and otherwise $\beta=\alpha$
, see Fig.(\ref{tconf}).
In $|\Psi[u]\rangle$, each $|\beta\rangle$ carries a factor 
$u_{lj}u_{in}$.
Eq. (\ref{z1}) indicates an additional option to get  $\alpha_{ij}$,
from $\alpha_{ij}$.

Taking the overlap with $\langle\Psi[u]|$, we get the matrix element which is expressed in
terms of Eq.(\ref{csisj222}). $\alpha_{ij}$ represents the Ket. A possible definition 
of the bonds of the Ket is the forward bonds  of the sites on sublattice $A$.

\subsection{Matrix element of single hole hopping term}
\label{hopterm}
\begin{figure}
\centerline{ \epsfysize=2.5cm
\epsfbox{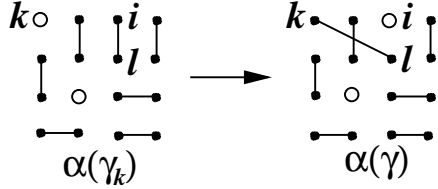}}
\caption{ The  operator $f^\dagger_i f_k (a^\dagger_k a_i +b^\dagger_k b_i)$ turn
a hole pair configuration $\gamma_k$,  $\alpha(\gamma_k)$ (left), with
 $k \in \gamma_k$, $(i, l) \in \alpha(\gamma_k)$,
to the right configuration $ \gamma$, $\alpha(\gamma)$ 
with $i \in \gamma$ and $(k,l) \in \alpha(\gamma)$.
In $f^\dagger_i f_k (a^\dagger_k a_i +b^\dagger_k b_i)|\Psi\langle$, 
this configuration has the coefficients $u_{il}$, and $\det\!V(\gamma_k)$.}
\label{phopp}  	 	 
\end{figure}

Fig.\ref{phopp}  describes the effect of a single hole hoping term on a hole-pair configuration.
Using definition (\ref{ooooo}), for $i, k\in A$
\begin{equation}
f^\dagger_i f_k (a^\dagger_k a_i +b^\dagger_k b_i) \left(\gamma, \Lambda_{\gamma}\right) =
\left\{
\begin{array}{ll}
\frac{\det\!{V\!\left(\gamma_k\right)}}
{\det\!V\!\left(\gamma\right)}
\frac{u_{il}}{u_{kl}}
s_{ki}
 &\mbox{if } i \in \gamma,\;k \not\in \gamma  \\
0 &  \mbox{otherwise}
\end{array}\right.
\label{hme}
\end{equation}
where $i\not\in\gamma_k$, $k\in \gamma_k$, and otherwise $\gamma_k=\gamma$;
$(k,l)\in \Lambda_{\gamma}$ is the forward bond  of
$k$ (i.e. originated in the Ket);
and $s_{ki}=\pm 1$  comes from reordering the 
fermion operators. Relation (\ref{hme}) is simplified using\cite{thesis} 
\begin{equation}
\det\!V\!\left(\gamma_k\right) s_{ki}=
\det\!V\left(\gamma, i\rightarrow k \right) 
\label{sshopi}
\end{equation}
with 
\begin{equation}
V\left(\gamma, i\rightarrow k \right)_{rp}=\left\{
\begin{array}{ll}
v(k, j_p)  & i_r = i \\
V(\gamma)_{rp} & \mbox{ otherwise}
\end{array}
\right.
\label{ahtv}
\end{equation}

\subsection{Matrix elements of the double hopping terms}
\label{resddd}
For $u_r>0\;\forall r$ we express 
$\langle\Delta^{\dagger}_{kl} \Delta_{ij}\rangle$, where $\Delta$ is given in Eq. (\ref{dalte}).
$\Delta^{\dagger}_{kl}$ creates a singlet bond. 
$\Delta_{ij}$ creates a pair of holes. With the results of App.\ref{c222}, for $i\in A, \; j\in B$
 \begin{eqnarray}
\Delta^{\dagger}_{kl} \Delta_{ij}\left(\gamma,\Lambda_{\gamma}\right) = \left\{
\begin{array}{ll}
\frac{s}{2u_{kl}}
\frac{\det \!V(\gamma_a)}
{\det \!V(\gamma)}
\left(\sum\limits_{
\scriptstyle (r,n)\in \alpha(\gamma) \atop 
\scriptstyle r\neq k\!\!}
\frac{u_{rj}u_{in}}{u_{rn}}
+2u_{ij}\right) &  \; 
\begin{array}{lll}
\mbox{ if }(k,l)\in
 \alpha(\gamma), \\ i\in \gamma  
 \mbox{ if } i \neq k,  \\ j \in \gamma  \mbox{ if } j \neq l. 
\end{array} \\
\\ 0 &\; \mbox{ otherwise}
\end{array}\right.
\label{mea}
\end{eqnarray}
Where $s=-1$ if $i =k$ exclusive-or $j =l$ and 1 otherwise,
$V(\gamma_a) \equiv V(\gamma, i\rightarrow k, j \rightarrow l)$ is defined like 
Eq. (\ref{ahtv}), with a possible replacement 
 of a row {\em and} a column,
and $\alpha(\gamma)$ is the set of forward bonds of A sub-lattice sites in $\Lambda(\gamma)$.

\section{The Gutzwiller  Approximation.}
\label{largen}

The  Gutzwiller  Approximation amounts to dropping the projector ${\cal P}(x)$ in definition (\ref{states2}) and setting 
$|\Psi [u,v;x]\rangle \rightarrow |\bar{\psi}[yu,zv]\rangle=|yu\rangle \otimes |zv\rangle$.
The constants $y=y(u)$ and $z=z(v)$ are determined by global constraint equations 
\begin{eqnarray}
\langle a^{\dagger}_i a_i \rangle &=& \langle b^{\dagger}_i b_i \rangle =  (1- x)/2,
\label{mfeq}\\
\langle f^{\dagger}_i f_i \rangle &=& x
\label{ffmfeq}
\end{eqnarray}
for $y$, $z$ respectively.  
In this section $\langle...\rangle\equiv\langle\bar{\psi}|...|\bar{\psi}\rangle/\langle\bar{\psi}|\bar{\psi}\rangle$.

$|yu\rangle$ is a  Schwinger bosons mean field wave function\cite{Book}, on which we preform the 
Marshall transformation\cite{Book}:
$a_j \rightarrow - b_j, \;b_j \rightarrow a_j,\; j \in B$.
 Hence
\begin{eqnarray}
|yu\rangle \rightarrow
\exp\left(
y\sum_{ij}u_{ij}
(a^{\dagger}_i a^{\dagger}_j + b^{\dagger}_ib^{\dagger}_j)\right)|0\rangle.
\label{sbmfs}
\end{eqnarray}
Operators are transformed  accordingly, for example 
$S^-_ j \rightarrow - { a}^{\dagger}_ j  { b}_ j$ for $j \in B$.

From  Eqs. (\ref{mfeq}) and  (\ref{ffmfeq}) 
\begin{equation}
\langle  S^+_i S^-_ i (1-f^{\dagger}_if_i)^2\rangle=
\langle n^ a_i(1+n^b_i)\rangle 
\langle(1-n^f_i)^2\rangle=
\frac{1- x}{2}\left(1+\frac{1- x}{2}\right)\left(1-x\right)^2
\label{sparefac}
\end{equation}
Whereas 
$\langle\Psi|  S^+_ i  S^-_ i (1-f^{\dagger}_if_i)^2|\Psi\rangle/\langle\Psi|\Psi\rangle = (1-x)/2$.
Thus we use 
\begin{equation}
(1-x/3)^{-1} \langle S^+_i  S^-_ j  \rangle
\label{mffh}
\end{equation}
as the GA  for the long range magnetic correlations and $m_0$  in the doped RVB state. {\em Empirically}
we omit the $(1-x/3)^{-1}$ factor in the estimates of magnetic energy.

Using the extended Wick theorem \cite{maxim}, for $i \in A$
\begin{eqnarray}
\langle  S^+_ i  S^-_ j\rangle= \left\{
\begin{array}{ll} 
-\langle  a^{\dagger}_ i   b_ i a^{\dagger}_ j b_ j\rangle = 
- \langle a^{\dagger}_i a^{\dagger}_j \rangle 
\langle b_i b_j \rangle \equiv - \rho_{ij}^2 \;\;j\in B \\
\langle  a^{\dagger}_ i   b_ i  b^{\dagger}_ j a_ j\rangle =  \langle a^{\dagger}_i a_j \rangle 
\langle b^{\dagger}_j  b_i \rangle + \delta_{ij}/2 \equiv \sigma_{ij}^2+ \delta_{ij}/2\;\;j\in A
\end{array}
\right. 
\label{deqd}
\end{eqnarray}
where we used, for example, $\langle a^{\dagger}_i  b_j\rangle = \langle a^{\dagger}_i  b^{\dagger}_j\rangle =0$.

Expanding 
$u_{\bf k}=\sum_j e^{i{\bf k}j}u_{0j}$, 
$\rho_{\bf k} = \sum_j e^{i{\bf k}j} \rho_{0j}$ and a similar expression for  $\sigma_{\bf k}$\cite{maxim},
\begin{equation}
\rho_{\bf k}  =
\frac{yu_{\bf k}}{1-y^2u^2_{\bf k}} 
\label{deq}
\end{equation}
and
\begin{equation}
\sigma_{\bf k}  =
\frac{y^2u^2_{\bf k}}{1-y^2u^2_{\bf k}}.
\label{sigq}
\end{equation}
The constraint equation (\ref{mfeq}) becomes
\begin{equation}
\frac{y^2}{L^2}\sum_{\bf k}\frac{u^2_{\bf k}}{1-y^2u^2_{\bf k}}=\frac{1-x}{2}.
\label{kcoeq}
\end{equation}
We consider  three cases:

\subsection{Exponential bond amplitude}
We calculate spin-spin correlation function for
\label{exuyy}
\begin{equation}
u(r)=
\left\{ \begin{array}{ll}
 \exp\left(-r/\xi \right)  & {\bf r}\; \mbox{bipartite}\\
0   & \mbox{otherwise} 
\end{array} \right. 
\label{adef}
\end{equation}
with  $\xi >> 1$. 
\begin{equation}
u_{{\bf k}}= \left\{	\begin{array}{ll} 
\int d^2{\bf r}  \exp\left(-r/\xi \right) \; \exp{\left(-i {\bf k}\cdot 
{\bf r} \right)} = 2 \pi \xi^{-1}  /\left(k^2+\xi^{-2}\right)^{\frac{3}{2}} \;\;\; & {\bf k} \in \mbox{MBZ}\\
u_{{\bf k}} = - u_{{\bf k-\pi}} & \mbox{otherwise}\\
 \end{array} \right.
\end{equation}
where MBZ= magnetic Brillouin zone.  Eq.(\ref{kcoeq}) becomes 
\begin{equation}
\frac{2 y^2}{ \xi^2}\int_{MBZ} \frac{d^2\!k}{\left(k^2+\xi^{-2}\right)^3 - \left(2 \pi  \xi^{-1} y\right)^2} = 
\frac{\pi y^2}{\xi^2}  \int_{\xi^{-2}}^{\infty} \frac{dk}{k^3 - \left(2 \pi \xi^{-1} y\right)^2}=
\frac{1 -  x}{2}
\label{cex}
\end{equation}  
where we multiplied the left side in $2$ to account for the integration over the complete Brillouin Zone. 
In all our calculations we took the continuum  limit (lattice constant$\rightarrow$0), 
where the upper bound of the integration $\rightarrow \infty$. 
This approximation works very well for slow decaying bond amplitude \cite{thesis}. 
 
Eq.(\ref{cex}) gives \cite{gradstein} 
\begin{eqnarray}
&&\frac{1}{3 \left(2 \pi \right)^{\frac{1}{3}}}\left(\frac{y}{\xi}\right)^{\frac{2}{3}}
\left\{-\pi\sqrt{3}  
-  2\sqrt{3} \arctan{ \left[ \frac{1}{\sqrt{3}}\left(1+\left(\frac{2}{\pi^2 y^2\xi^4}\right)^{\frac{1}{3}}\right)\right]}  
\right. \nonumber \\ 
 &+& \left.
 \ln{\left[ 4 \left(\frac{\pi y}{ \xi} \right)^{\frac{4}{3}} + 2 \left(\frac{2\pi^2 y^2}{\xi^8}\right)^{\frac{1}{3}} + 
\frac{2^{\frac{2}{3}}}{ \xi^4} \right]}
- 2 \ln{\left[ \frac{2^{\frac{1}{3}}}{ \xi^2} - 2 \left(\frac{\pi y}{ \xi} \right)^{\frac{2}{3}}\right]}
 \right\} =\frac{1 -  x}{2}
\label{ix}
\end{eqnarray}
The argument of the last logarithm 
has to be sufficiently  close to zero for Eq.(\ref{ix}) to be satisfied. Therefore 
\begin{equation}
2^{\frac{1}{3}} \xi^{-2} - 2\left(\pi y \xi^{-1}\right)^{\frac{2}{3}} \approx 0 \Longrightarrow 
y \approx \frac{1}{2\pi\xi^{2}}.
\label{az}
\end{equation}
Hence we can neglect in the left side of Eq.(\ref{ix}) all terms but the last log.
Consequently
\begin{equation}
y^2 \cong \frac{1}{(2\pi)^2\xi^{4}}
\left\{1 -  \frac{3\xi^{2}}{2^{\frac{1}{3}}} \exp{\left[-\left(1- x \right)
\frac{3\pi \xi^{2}}{2}\right]}\right\} \equiv \frac{1}{(2\pi)^2\xi^{4}}\cal{D}
\label{x2}
\end{equation}

Eq.(\ref{deqd}) becomes
\begin{eqnarray}
\sigma_{\bf r}=2\int_{MBZ}  d^2{\bf k}\sigma_{\bf k}\exp{\left(i{\bf{k\cdot r}}\right)}=
\frac{ {\cal D}}{2 \pi \xi^{6}}\int  dk\frac{kJ_0\left(kr\right)}
{\left({\bf k}^2 + \xi^{-2}\right)^3 - \xi^{-6}\cal{D}}
\label{sq2}
\end{eqnarray}
where $J_0$ is the 
Bessel function\cite{gradstein}.
Since the integrand in Eq.(\ref{sq2}) vanishes as $k \rightarrow 0$, 
we can replace $J_0$  with its  approximation for
$k r\gg 0$. Expanding the denominator to first order in $k^2$ 
\begin{eqnarray}
\sigma_{\bf r}\cong
\frac{ {\cal D}}{6 \pi \sqrt{r\pi} \xi^{2}}\int_0^{ \infty} dk\frac{\sqrt{k}}
{k^2 +  a^2}\left[\cos\left(kr\right) +\sin\left(kr\right)\right] \equiv \frac{ {\cal D}}{6 \pi \sqrt{r\pi} \xi^{2}}Y_0
\label{sq4}
\end{eqnarray}
where $a^2=(1-{\cal D})/(3\xi^{2})$. In the definition 
\begin{eqnarray*}
Y_1\equiv \int_0^{ \infty} dk\frac{\sqrt{k}}
{k^2 +  a^2}\exp{\left(ikr\right)}
\end{eqnarray*}
 $Y_0= ReY_1+ImY_1$.
Let us consider the Integral
\begin{eqnarray*}
Y_2=\oint dz\frac{\sqrt{z}}
{z^2 +  a^2}\exp{\left(izr\right)} = \int_{e^{i\pi}\infty}^{ \infty} dk\frac{\sqrt{k}}
{k^2 +  a^2}\exp{\left(ikr\right)}
\end{eqnarray*}
where the close contour  encircles the upper half of the complex plane.
The part of the contour along the negative real axis is
\begin{eqnarray*}
\int_{e^{i\pi}\infty}^{0} dk\frac{\sqrt{k}}
{k^2 +  a^2}\exp{\left(ikr\right)} =  i \int_0^{\infty}dk'\frac{\sqrt{k'}}
{k'^2 +  a^2}\exp{\left(-ik'r\right)} = iY^{\star}_1
\end{eqnarray*}
where we substituted  $k'=e^{-i\pi}k$. Hence $Y_2=Y_1+iY^{\star}_1=(1+i)(ReY_1+ImY_1)=(1+i)Y_0$.
Using the residue method for $Y_2$:
\begin{equation}
\sigma_{\bf r} \propto \frac{\exp\left[-r\sqrt{(1-{\cal D})/(3\xi^{2})}\right]}{\sqrt{r}},
\label{sir}
\end{equation}
and with Eq. (\ref{x2}) we find for the correlation length of the spin-spin correlation function, $\xi_{ex}$ :
\begin{equation}
\xi_{ex} \approx \sqrt{\frac{3\xi^{2}}{4 \left(1-\cal{D}\right)}} =
 2^{\frac{1}{3}} \exp{\left[\left(1-x \right) \frac{3 \pi}{2} \xi^2\right]}
\end{equation}

\subsection{Gaussian  bond amplitude}
\label{andel}
We calculate spin-spin correlation function for
\begin{eqnarray*}
u_{{\bf r}}=  \left\{ \begin{array}{ll}
\exp\left(- \mu r^2  \right) & {\bf r}\; \mbox{bipartite}\\
0   & \mbox{otherwise} 
\end{array} \right. 
\end{eqnarray*}
For ${\bf k} \in$ MBZ
\begin{equation}
u_{{\bf k}} = \frac{\pi}{\mu}\exp{\left(-\frac{k^2}{4\mu}\right)}
\label{gusu}
\end{equation}
Eq.(\ref{kcoeq}) is 
\begin{equation}
 \frac{\mu}{2\pi}\int_{0}^{1} \frac{dt}{\frac{\mu^2}{\pi^2y^2}-t} = 
\frac{1 -  x}{4}
\label{ceg}
\end{equation}
with the solution 
\begin{equation}
y^2 =  \frac{\mu^2}{\pi^2}\left\{1 - \exp{\left[-\left(1 - x\right) 
\frac{\pi}{2\mu}\right]}\right\} \equiv \frac{\mu^2}{\pi^2}{\cal G}
\label{xgxx}
\end{equation}
Calculation of $\sigma_{\bf r}$ is identical to the exponential case.
Substituting  in Eq. (\ref{sq4}) $a^2=2\mu(1-{\cal G})$ 
\begin{eqnarray}
\sigma_{\bf r} \propto
\frac{\exp{\left(-r \sqrt{2\mu\left(1-{\cal G}\right)}\right)}}{\sqrt{r}}
\label{frexp}
\end{eqnarray}
and hence the spin-spin correlation function decays exponentially with correlation length
\begin{equation}
\xi_g=\frac{1}{\sqrt{8 \mu \left(1-\cal{G}\right)}} =
\frac{1}{\sqrt{8\mu}} \exp{\left[\left(1-x \right) \frac{\pi}{4\mu}\right]}
\end{equation}

\subsection{Power Law  bond amplitude}
\label{pluan}
For the bond amplitude
\begin{equation}
u_{\bf r}=  \left\{ \begin{array}{ll}
 \frac{\epsilon^3}{\left(r^2+\epsilon^2\right)^{\frac{3}{2}}}& {\bf r}\; \mbox{bipartite}\\
0   & \mbox{otherwise} 
\end{array} \right. 
\label{plu}
\end{equation}
we show, in the continuum limit, that for   $0<\epsilon < \epsilon_0$,  $S(\pi,\pi)$ 
is finite and hence $m_0=0$.

Calculations of the GA 
on lattices of size $L\leq 512$, show that for  any $\epsilon$, the spin-spin correlation function calculated with
 function (\ref{plu}), decayes slower than with $u=1/r^3$. This suggests that $m_0=0$  for   $u=1/r^3$.

\begin{eqnarray*}
S(\pi,\pi)= \sum_ j |\langle S^+_{{\bf r}_j}S^-_0\rangle| = \sum_j \sigma^2_{{\bf r}_j}+ 
\rho^2_{{\bf r}_j} +\frac{1}{2} = \frac{1}{L^2} \sum_{{\bf q}} 
\left(\sigma^2_{{\bf q}} + \rho^2_{{\bf q}}\right) +\frac{1}{2}.
\end{eqnarray*}
where we used Eq.(\ref{deqd}). For ${\bf k} \in$ MBZ \cite{gradstein}
\begin{equation}
u_{{\bf k}} = 
2\pi \epsilon^2 e^{-\epsilon k}
\label{kplu}
\end{equation}
From Eqs.(\ref{deq}) and (\ref{sigq}) $\rho_{{\bf k}}\stackrel{k\rightarrow \infty}{\longrightarrow} e^{-\epsilon k}$, 
and $\sigma_{{\bf k}}\stackrel{k\rightarrow \infty}{\longrightarrow} e^{-2\epsilon k}$. Hence $S(\pi,\pi)$ might diverges 
only if there is $k_0$, such that $\left(1-ae^{-2\epsilon k}\right)_{k=k_0}=0$, where
$a=a(\epsilon)=\left(2\pi y \epsilon^2\right)^2$. Therefore if $a<1$,  $S(\pi,\pi)$ is  finite.

Eq.(\ref{kcoeq}) for $y$ is 
\begin{equation}
\frac{a}{\pi}\int_0^{\infty}dk\frac{ k e^{-2\epsilon k}}{1-ae^{-2\epsilon k}} = 
\frac{a}{\pi}\int_0^{\infty}dk\frac{ k }{e^{2\epsilon k}-a}=\frac{1-x}{2}
\label{coeqpl}
\end{equation}
which becomes  \cite{gradstein}
\begin{equation}
\sum\limits_{p=1}^{\infty}\frac{a^p}{p^2}=\pi\epsilon^2(1-x)
\label{sumpl}
\end{equation}
The right side of Eq.(\ref{sumpl}) is $y$ independent, and increases with $\epsilon$, hence
$a$ increases with $\epsilon$. Therefore $a(\epsilon_0) = 1$.
For $a=1$, the left side of Eq.(\ref{sumpl}) is $\pi^2/6$ and 
\begin{equation}
\epsilon_0 = \sqrt{\frac{\pi}{6(1-x)}},
\label{ep0}
\end{equation}
and for $\epsilon < \epsilon_0$, $S(\pi,\pi)$ is finite and hence $m_0=0$.


\begin{table}
\begin{tabular}{|c|c|c|c|} \hline
state                    &$E_0(x=0)$ [J/bond]  &$E_0$ parameters   & $m^2_0(x=0)$ \\ \hline
$u_{ex}$, Eq. (\protect{\ref{uex}}) &$-0.335\pm 0.0005$& $\xi^{-1}=0.17\pm0.005$        & $0.125\pm0.005$ \\\hline   
$u_g$, Eq. (\protect{\ref{ug}})     &$-0.335\pm 0.0005$  & $Q=0.014\pm0.001$& $0.12\pm0.005$ \\ \hline
$u_p$, Eq. (\protect{\ref{uppl}}) &$-0.335\pm 0.0005$   & $p=2.7\pm0.05$& $0.105\pm0.005$ \\ \hline
$u_{MF}$&  $-0.3344\pm 0.0002$  &                       & $0.087\pm0.005$      \\ \hline
\end{tabular}
\caption{Minimal magnetic energy ($E_0$), optimal choice of parameters, and square of the
ground state magnetization ($m^2_0$),
 for various  bond amplitudes in the undoped, $x=0$, case. 
The size of the Lattice is $L=40$.
 $u_{MF}$ is derived from the Schwinger bosons mean field theory of ${\cal H}^J$.
The ground state parameters of the Heisenberg model on an $L=40$ lattice are : E(ground state)= 0.3347 J/bond 
and $m_0^2$(ground state)=0.109{\protect\cite{Sandvik}}.}      
\label{paraatmin}
\end{table}

\end{document}